\documentclass[twocolumn,tighten]{aastex631}
\accepted{to AAS journals on 12/29/2021}
\usepackage{graphicx}
\usepackage{natbib}
\usepackage{latexsym}
\usepackage{amssymb}
\usepackage{longtable}
\usepackage{amsmath}
\usepackage{url}
\def\msun{\ifmmode {\rm\,M_\odot}\else ${\rm\,M_\odot}$\fi}
\def\Msun{\ifmmode {\rm\,\it{M_\odot}}\else ${\rm\,M_\odot}$\fi}
\def\lsun{\ifmmode {\rm\,L_\odot}\else ${\rm\,L_\odot}$\fi}
\def\Lsun{\ifmmode {\rm\,\it{L_\odot}}\else ${\rm\,L_\odot}$\fi}
\def\rsun{\ifmmode {\rm\,R_\odot}\else ${\rm\,R_\odot}$\fi}
\def\Rsun{\ifmmode {\rm\,\it{R_\odot}}\else ${\rm\,R_\odot}$\fi}
\def\Tsun{\ifmmode {\rm\,T_\odot}\else ${\rm\,T_\odot}$\fi}
\def\arcsec{\ifmmode {^{\prime\prime}}\else $^{\prime\prime}$\fi}
\def\asec{\ifmmode {^{\prime\prime}}\else $^{\prime\prime}$\fi}
\def\arcmin{\ifmmode {^{\prime}}\else $^{\prime}$\fi}
\def\amin{\ifmmode {^{\prime}}\else $^{\prime}$\fi}
\def\simlt{\mathrel{\spose{\lower 3pt\hbox{$\mathchar"218$}}
     \raise 2.0pt\hbox{$\mathchar"13C$}}}
\def\simgt{\mathrel{\spose{\lower 3pt\hbox{$\mathchar"218$}}
\     \raise 2.0pt\hbox{$\mathchar"13E$}}}

\begin{document}

\author{P. Wilson Cauley}
\affiliation{Laboratory for Atmospheric and Space Physics, University of Colorado Boulder, Boulder, CO 80303}

\author{John P. Ahlers}
\altaffiliation{NASA Postdoctoral Program Fellow}
\affiliation{Exoplanets and Stellar Astrophysics Laboratory, Code 667, NASA Goddard Space Flight Center (GSFC), Greenbelt, MD 20771, USA}
\affiliation{GSFC Sellers Exoplanet Environments Collaboration}

\correspondingauthor{P. Wilson Cauley}
\email{pwcauley@gmail.com}

\title{The effects of stellar gravity darkening on high-resolution transmission spectra}

\begin{abstract}

High-resolution transmission spectroscopy is a powerful method for probing the extended atmospheres of short-period exoplanets. With the advancement of ultra-stable echelle spectrographs and the advent of 30-meter class telescopes on the horizon, even minor observational and physical effects will become important when modeling atmospheric absorption of atomic species. In this work we demonstrate how the non-uniform temperature across the surface of a fast rotating star, i.e., gravity darkening, can affect the observed transmission spectrum in a handful of atomic transitions commonly observed in short-period exoplanet atmospheres. We simulate transits of the ultra-hot Jupiters KELT-9 b and HAT-P-70 b but our results are applicable to all short-period gas giants transiting rapidly rotating stars. In general, we find that gravity darkening has a small effect on the average transmission spectrum but can change the shape of the absorption light curve, similar to the effect observed in broadband photometric transits. While the magnitude of gravity darkening effects are on the same order as the noise in transmission spectra observed with 10-meter class telescopes, future high-quality spectroscopic light curves for individual atomic absorption lines collected with 30-meter class telescopes will need to account for this effect.  

\end{abstract}


\section{INTRODUCTION}
\label{sec:intro}

Transmission spectroscopy is one of the primary tools for measuring the properties of exoplanet atmospheres. Different layers of planetary atmospheres can be observed by exploiting the varying populations of molecular and atomic transitions across a range of temperatures and pressures. As a result, transmission spectroscopy has revealed detailed physical and chemical information about planetary atmospheres, from photospheres \citep[][]{deming13,kreidberg14,sing16,weaver21} to thermospheres \citep[][]{redfield08,wyttenbach15,casasayas17,salz18,yan18,cauley19,seidel20a,hoeijmakers19,cauley21} and all the way out to unbound exospheres \citep[e.g.,][]{bourrier13,ehrenreich15,allart18,bourrier18,zhang21}. 

As instruments become more sensitive and processing techniques more advanced, it is becoming important to understand all of the physical and observational factors which contribute to transmission spectrum signals. This includes the possible contributions of stellar activity \citep[][]{rackham18,cauley18,salz18}, stellar center-to-limb variations (CLVs) and the planet's Doppler shadow \citep[][]{czesla15,yan15,yan17}, stellar pulsations \citep{yan21,cauley21}, and changes in the transit chord due to orbital precession \citep[][]{herman18,johnson18,watanabe20,szabo20}. For high-resolution transmission spectra these effects can alter the observed line profile morphologies, as well as the shape of light curves for individual atomic absorption lines. 

One aspect of transmission spectroscopy contamination that is relevant to planets transiting rapidly rotating stars is the non-spherical shape and non-uniform surface temperature of their host stars, the result of gravity darkening from the star's high spin rate \citep{barnes09}. For transiting planets with large spin-orbit misalignment it is possible to derive the star's gravity darkened parameters, which includes the true inclination of the stellar spin axis in the plane of the sky, the ratio of the polar to equatorial radii, i.e. the star's oblateness, and the surface temperature as a function of latitude \citep[e.g.,][]{barnes11,masuda15,zhou19,ahlers20,ahlers20a,garai21}. 

Recently, a number of ultra-hot gas giants transiting rapidly rotating A-type stars have been the focus of detailed transmission spectroscopy campaigns due to their novel atmospheric chemistry and high signal-to-noise absorption signatures \citep{yan18,hoeijmakers19,cauley19,casasayas19,nugroho20a,hoeijmakers20,stangret20,rainer21,yan21,cauley21,bello21}. Given the intense interest in these systems, and the likely discovery of more such objects in the \textit{TESS} data, it is important to understand the magnitude and details of how gravity darkening can affect the observed transmission spectrum. 

In this paper we explore this question for two ultra-hot Jupiter systems, KELT-9 b \citep{gaudi17} and HAT-P-70 b \citep{zhou19}, by comparing transmission spectroscopy models of strong atomic transitions for the gravity-darkened and non-gravity darkened cases. KELT-9 b and HAT-P-70 b are useful cases to examine due to the large spin-orbit misalignment in both systems, which results in the planets transiting chords of non-uniform temperature. For both systems we compare the transit light curves and the transmission spectrum line profile morphology, for each atomic absorber, in the case of a gravity-darkened and uniform temperature host star. In \autoref{sec:stellpars} we derive the gravity-darkened parameters of both systems using \textit{TESS} photometry. We then describe the transmission spectrum models in \autoref{sec:tspecs} and show the model comparisons in \autoref{sec:results}. We finish by summarizing our main findings in \autoref{sec:conclusion}.        

%For planets with zero orbital obliquity \citep[e.g., KELT-20 b;][]{lund17} the planet will transit a chord with uniform temperature and thus produce a very similar transmission signal compared with the non-gravity darkened case.

\section{GRAVITY-DARKENED STELLAR PARAMETERS}
\label{sec:stellpars}

In our analysis of the KELT-9 and HAT-P-70 systems, we account for the rapid rotation observed in both host stars. Rapid rotation induces two effects on a star that change its total flux output. First, the star flattens into an oblate shape, with an equatorial radius larger than a polar radius. Second, the star's effective temperature varies in a pole-to-equator gradient, with hot, bright poles and a cooler, dim equator. 

We model these effects in both stars using the rotating star model derived in \citet{espinosalara2011}. Our transit fitting model accounts for stellar flattening and gravity darkening following the same path established by \citet{barnes09}, but with the von Zeipel theorem \citep{von1924radiative} replaced by the Roche model of a rotating star. We model each system's primary transit using Equations 6 and 7 of \citet{barnes09}, but we use updated expressions for the star's flattening and temperature gradient due to rotation. We calculate each host star's radius and effective temperature using Equations 30 and 31, respectively, from \citet{espinosalara2011}. These equations replace the expressions in \citet{barnes09} for stellar radius and temperature (Equations 12 and 9, respectively). Aside from these changes, we follow the transit-fitting method derived in \citet{barnes09} and applied in several works \citep[e.g.,][]{barnes11,ahlers2014spin,ahlers20,ahlers20a}.

\begin{figure*}[htbp]
   \centering
   \begin{tabular}{rl}
   \includegraphics[width=.47\textwidth]{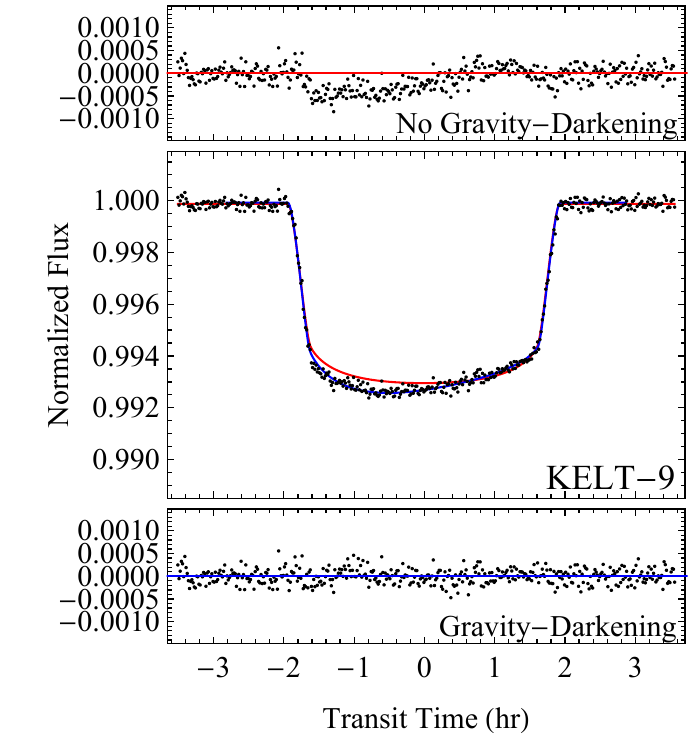} & \includegraphics[width=.47\textwidth]{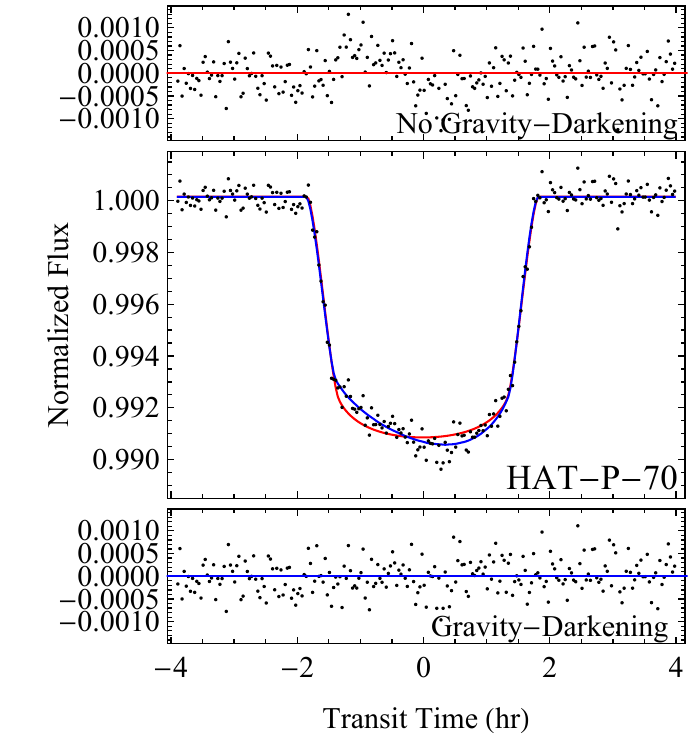} 
   \end{tabular}
   \figcaption{\emph{TESS} photometry and best-fit results for KELT-9 and HAT-P-70. Both transits are visibly asymmetric due to stellar gravity darkening. The gravity-darkened model (blue) yields an improved fit to both datasets versus the spherical star model (red).
\label{fig:hatp70fit}}
\end{figure*}

\begin{figure}[htbp]
   \centering
   \includegraphics[scale=.5,clip,trim=25mm 25mm 15mm 25mm,angle=0]{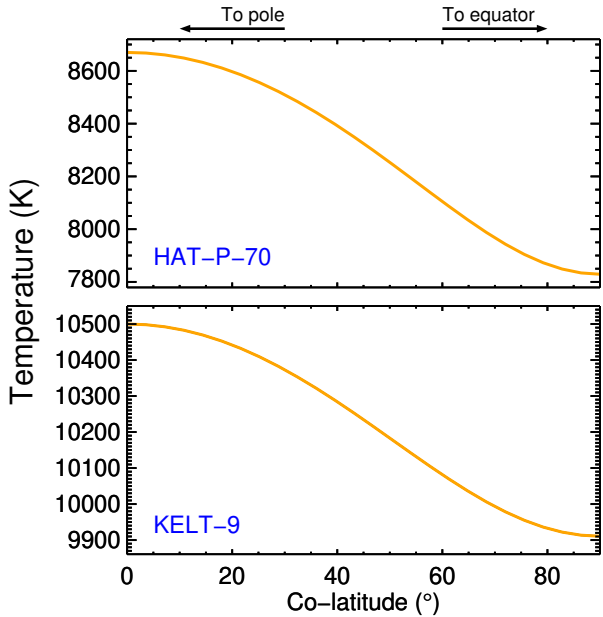}
   \figcaption{Co-latitude versus temperature for HAT-P-70 (top) and KELT-9 (bottom). HAT-P-70's polar temperature is almost 850 K greater than its equatorial temperature. For KELT-9 the difference is approximately 700 K. 
\label{fig:tcolat}}
\end{figure}

For both stars, we measure gravity-darkened stellar parameters from \emph{TESS} photometry using the same approach as \citet{ahlers20}. We use the light curve fitting package {\tt transitfitter} \citep{barnes09}, which accounts for gravity darkening in primary transit photometry and which has been updated to use the rotating star model derived in \citet{espinosalara2011}. Taking advantage of the asymmetry in the primary transit of both systems, we measure each planet's projected alignment and the stellar inclination, which together yield the planet's true spin-orbit angle. We simultaneously fit for the stellar rotation rate, oblateness, and temperature gradient, yielding a full description of each host star's gravity darkening. The best-fit parameters for each system are given in \autoref{tab:pars}. We show the best-fit models to the \emph{TESS} photometry in \autoref{fig:hatp70fit} and list relevant parameters of both systems in \autoref{tab:pars}. We show the best-fit temperature as a function of co-latitude for both planets in \autoref{fig:tcolat}.

Our measured values for HAT-P-70 differ slightly from the measured values in \citet{zhou19}. Specifically, our measured stellar inclination of $i_\star=35^{+11}_{-8}$ degrees does not overlap with their $1\sigma$ results of $i_\star=58.8^{+7.5}_{-4.8}$ degrees. This difference likely occurs due to the difference in our stellar models. \citet{zhou19} used the von Zeipel rotating star model \citep{von1924radiative} and fit the gravity-darkening exponent $\beta=0.242^{+0.026}_{-0.029}$, which serves as a scaling factor for the strength of the star's temperature gradient. In this work, we use the rotating star model derived in \citet{espinosalara2011}, which does not include a $\beta$ term. As a comparison, we find that the $\beta$ that best recreates the temperature gradient in the \citet{espinosalara2011} model is $\sim0.20$, which is significantly lower than the measured value in \citet{zhou19}. A higher $\beta$ can skew $i_\star$ values toward $90^\circ$ \citep{barnes09}. Therefore, our HAT-P-70 results likely differ from \citet{zhou19} because we model the star's temperature to vary less between equator and pole. We note that \citet{espinosalara2011} use two-dimensional rotating star models to find the best-fit $\beta$ values as a function of stellar oblateness. They find that $\beta \approx 0.22$ for HAT-P-70's oblateness value $f = 0.08$. Thus we recommend that future studies of HAT-P-70 adopt the slightly lower values of $\beta$ found here and in \citet{espinosalara2011} to better approximate HAT-P-70's temperature gradient.

We also find slightly different stellar parameters for KELT-9 b compared with those from \citet{ahlers20}. This is again likely the result of our implementation of the rotating star model from \citet{espinosalara2011} compared with the use of the von Zeipel theorem in \citet{ahlers20}. We find a slightly hotter, slightly less oblate star than \citet{ahlers20}. Our derived rotation period for KELT-9 is also longer, although the values from both studies are consistent at the $1\sigma$ level. 

\begin{deluxetable*}{llcccc}
%\tablewidth{1.99\textwidth}
\tablecaption{KELT-9 and HAT-P-70 system parameters\label{tab:pars}}
\tablehead{\colhead{}&\colhead{}&\multicolumn{2}{c}{KELT-9}&\multicolumn{2}{c}{HAT-P-70}\\
\colhead{Parameter}&\colhead{Description}&\colhead{Value}&\colhead{Reference}&\colhead{Value}&\colhead{Reference}}
\colnumbers
\tabletypesize{\scriptsize}
\startdata
\textbf{Uniform surface parameters:} & & & & & \\
Stellar mass & $M_\star$ ($M_\Sun$) & $2.52^{+0.25}_{-0.20}$ & 1 & $1.890^{+0.010}_{-0.013}$ & 2\\
Stellar radius & $R_\star$ ($R_\Sun$) & $2.362^{+0.075}_{-0.063}$ & 1 & $1.858^{+0.119}_{-0.091}$ & 2\\
Stellar surface gravity & log$g$ (cm $^{-2}$) & $4.093\pm0.14$ & 1 & $4.181^{+0.055}_{-0.063}$ & 2\\
Effective temperature & $T_\text{eff}$ (K) & $10,170\pm450$ & 1 & $8450^{+540}_{-690}$ & 2\\
Metallicity$^\dagger$ & [Fe/H] & $0.0$ & 1 & $0.0$ & 2\\
Stellar rotational velocity & $v$sin$i$ (km s$^{-1}$)& $111.4\pm1.3$ & 1 & $99.9\pm0.6$ & 2\\
& & & & & \\
\textbf{Gravity-darkened parameters:} & & & & & \\
Stellar equatorial radius & $R_\text{eq}$ ($R_\Sun$) & $2.38\pm0.03$ & 3 & $2.17\pm0.02$ & 3 \\
Stellar oblateness & $f$ & $0.057\pm0.019$ & 3 & $0.08^{+0.04}_{-0.03}$ & 3 \\
Stellar inclination$^\star$ & $i_\text{*}$ ($^\circ$) & $45\pm8$ & 3 & $35^{+11}_{-8}$ & 3 \\
Stellar rotation period & $P_\text{*}$ (hours) & $18^{+4}_{-4}$ & 3 & $14\pm3$ & 3 \\
Polar effective temperature & $T_\text{pole}$ (K) & $10500\pm400$ & 3 & $8670\pm300$ & 3 \\
Equatorial effective temperature & $T_\text{eq}$ (K) & $9910\pm400$ & 3 & $7830\pm290$ & 3 \\
& & & & &\\
\textbf{Planetary parameters:} & & & & & \\
Projected alignment angle & $\lambda$ ($^\circ$) & $-87\pm16$ & 3 & $113\pm11$ & 3 \\
Orbital inclination & $i$ ($^\circ$) & $88.1\pm0.4$ & 3 & $96.5\pm1.2$ & 3 \\
%Spin-orbit angle & $\psi$ ($^\circ$) & $87^{+9}_{-11}$ & 3 & $107^{+8}_{-6}$ & 3 \\
Orbital period & $P_\text{orb}$ (days) & $1.4811235\pm0.0000011$ & 1 & $2.74432452^{+0.00000068}_{-0.00000079}$ & 2\\
Semi-major axis & $a$ (AU) & $0.03462^{+0.00110}_{-0.00093}$ & 1 & $0.04739^{+0.00031}_{-0.00106}$ & 2\\
Planetary radius & $R_\text{p}$ ($R_\text{J}$) & $1.891^{+0.061}_{-0.093}$ & 1 & $1.87^{+0.15}_{-0.10}$ & 2\\
Transit duration & $T_{14}$ (days) & $0.1632\pm0.0005$ & 1 & $0.1450^{+0.0028}_{-0.0020}$ & 2\\
% System velocity & $\gamma$ & km s$^{-1}$ & $3.69 \pm 0.09$ & 3 \\
\\
\enddata
\tablerefs{$1=$ \citet{gaudi17}; $2=$ \citet{zhou19}; $3=$ this work.}
\tablenotetext{^\dagger}{The published metallicities for both stars are consistent with $[\text{Fe}/\text{H}]=0.0$.}
\tablenotetext{^\star}{Inclination measured from the line-of-sight.}
\end{deluxetable*}

\begin{figure*}[htbp]
   \centering
   \includegraphics[scale=.75,clip,trim=15mm 15mm 5mm 20mm,angle=0]{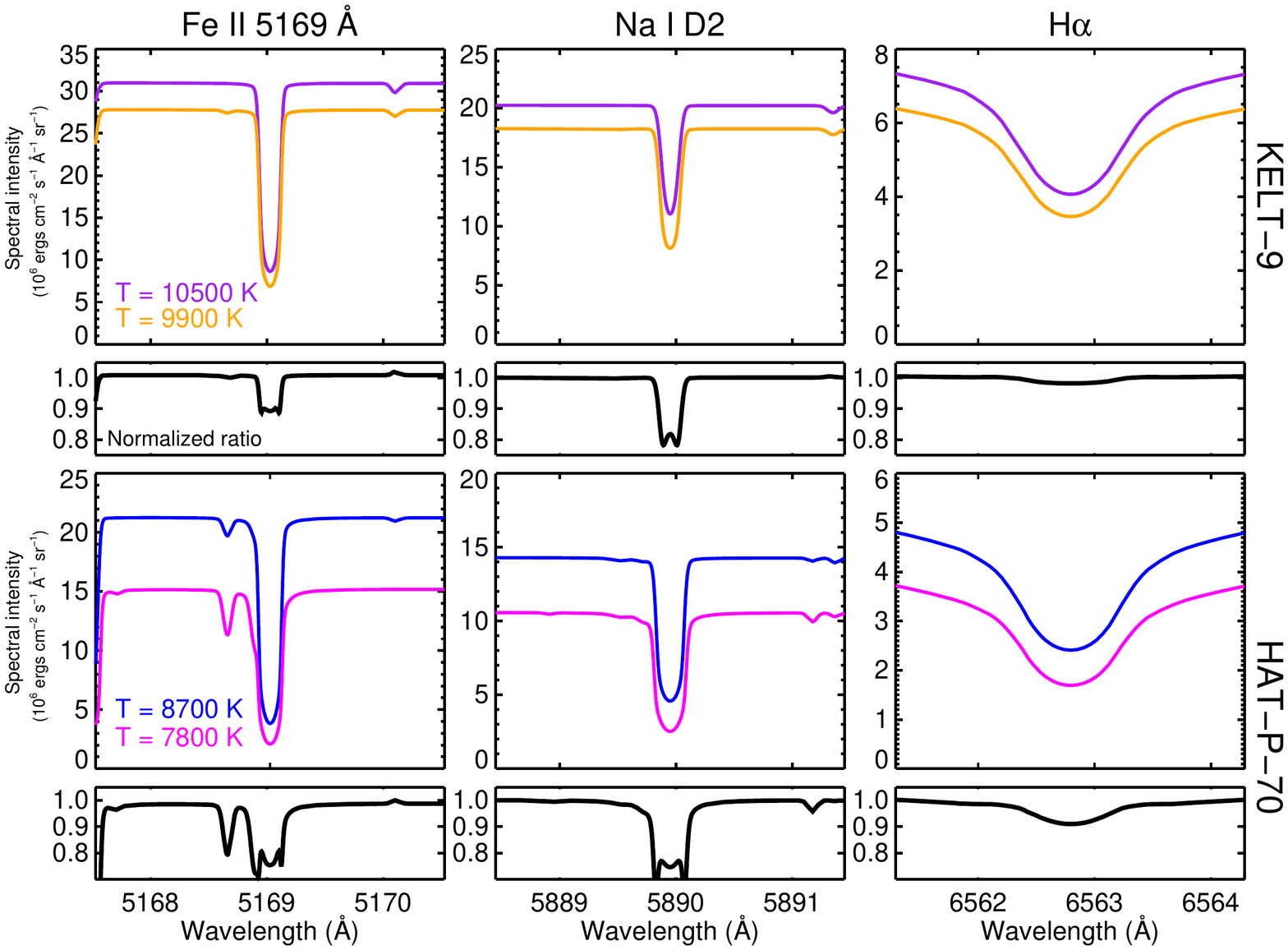}
   \figcaption{Comparison of disk-center ($\mu = 1.0$) synthetic spectra for the approximate polar and equatorial temperatures for KELT-9 (upper panels) and HAT-P-70 (lower panels). The intensity differences are a result of the drastically different temperatures. The bottom panels show the normalized ratio between the polar (hotter) and equatorial (cooler) spectra: differences of $\approx 10-30\%$ are apparent in the line cores of \ion{Fe}{2} 5169 \AA\ and \ion{Na}{1} D2 for both stars. 
\label{fig:scomps}}

\end{figure*}

Once we have the range of temperatures from the equator to the pole we generate synthetic spectra across that temperature range using \texttt{Spectroscopy Made Easy} \citep[\texttt{SME};][]{piskunov17}. For each atomic transition of interest we create stellar spectra in steps of $T=100$ K from the maximum polar temperature down to the equatorial temperature. At each temperature we also synthesize spectra at 15 different values of $\mu = \cos{\theta}$, where $\theta$ is the latitude on the stellar surface, to account for limb darkening and center-to-limb variations. In \autoref{fig:scomps} we show the disk-center ($\mu = 1$) spectra for the approximate polar and equatorial temperatures of each star. In general, the temperature gradient between pole and equator is not large enough to dramatically change the depth and shape of most atomic absorption lines. The most noticeable difference is the relative flux levels of the pole versus the equator, which will impact the shape of the spectroscopic light curve (see \autoref{sec:results}).  

We then populate a Cartesian grid of the stellar surface, where each pixel has dimensions $0.005 R_\text{eq} \times 0.005 R_\text{eq}$, by interpolating the synthesized spectra onto the specific temperature and $\mu$ value of each pixel, while also accounting for the local rotational velocity of the star. The result is a cube with two spatial axes and a wavelength axis containing a velocity-shifted and limb-darkened spectrum at each point on the oblate and inclined stellar surface. By summing over the stellar surface, i.e., the spatial dimensions of the cube, we generate the star's integrated spectrum which is used as the out-of-transit spectrum in the transmission calculations. We display the final surface temperature maps, accounting for stellar inclination, in \autoref{fig:tmaps}. We also show the planet transit chords which are calculated using \texttt{EXOFAST} \citep{eastman13}.

%\section{Modelling HAT-P-70's \emph{TESS} Photometry}

\section{TRANSMISSION SPECTRUM MODELS}
\label{sec:tspecs}

We use parameterized atmospheric models to calculate transmission spectra for KELT-9 b and HAT-P-70 b. The models are the same as those described in \citet{cauley19} and \citet{cauley21} and we briefly review them here. The planetary atmosphere is parameterized by the following values: the number density $n$ of the relevant atomic species, the radial distance $r$ above the optical planet radius, the thermal broadening $v_\text{t}$, Lorentzian broadening $v_\text{Lor}$, and rotational broadening $v_\text{rot}$. For simplicity we assume a uniform density atmosphere. We stress that the specific values of these parameters chosen for this study are not important since we are interested in the relative changes between the gravity darkened and non-gravity darkened cases. Thus we are more concerned with consistency of the planetary parameters between the gravity darkened and uniform temperature scenarios and less concerned about the accuracy of the atmospheric absorption compared with observations. No current detections of HAT-P-70 b's atmosphere have been published and we chose KELT-9 b's atmospheric parameters to roughly match the line depths and morphologies from the literature \citep{yan18,cauley19,hoeijmakers19}. 

We construct the planet's atmosphere on a Cartesian grid with the same spatial resolution as the stellar grid described in \autoref{sec:stellpars}. Once the 3D atmosphere is generated we collapse the 3D grid into a 2D grid of column densities at each $x-y$ pixel in the planet's atmosphere. We note that our assumptions of a spherically symmetric and uniform density atmosphere render 3D treatment unnecessary; however, the code is built to handle non-symmetric cases and 3D velocity fields so the initial 3D structure is critical in these cases. We then extinct the stellar spectrum at each pixel with non-zero column density, accounting for all sources of velocity shifts (e.g., the local planetary orbital and rotational velocities) and line broadening. We ignore multiple scattering events and refraction given their negligible impact on the transmission spectra of hot Jupiters \citep[e.g.,][]{betremieux16,robinson17}.

We simulate transmission spectra for three different atomic transitions that have been detected in a number of hot and ultra-hot Jupiter atmospheres: the $n=3 \rightarrow 2$ Balmer line H$\alpha$, the \ion{Na}{1} D2 5889.95 \AA\ doublet member, and an \ion{Fe}{2} transition at 5169.03 \AA. We chose this particular \ion{Fe}{2} line because it was one of the strongest \ion{Fe}{2} transitions detected in the atmospheres of KELT-9 b and KELT-20 b \citep{cauley19,casasayas19}, although our results are applicable to other \ion{Fe}{2} lines with similar temperature sensitivity. The atomic line parameters are sourced from the NIST Atomic Spectra Database\footnote{https://dx.doi.org/10.18434/T4W30F}. 

\begin{figure*}[htbp]
   \centering
   \includegraphics[scale=.67,clip,trim=5mm 15mm 5mm 80mm,angle=0]{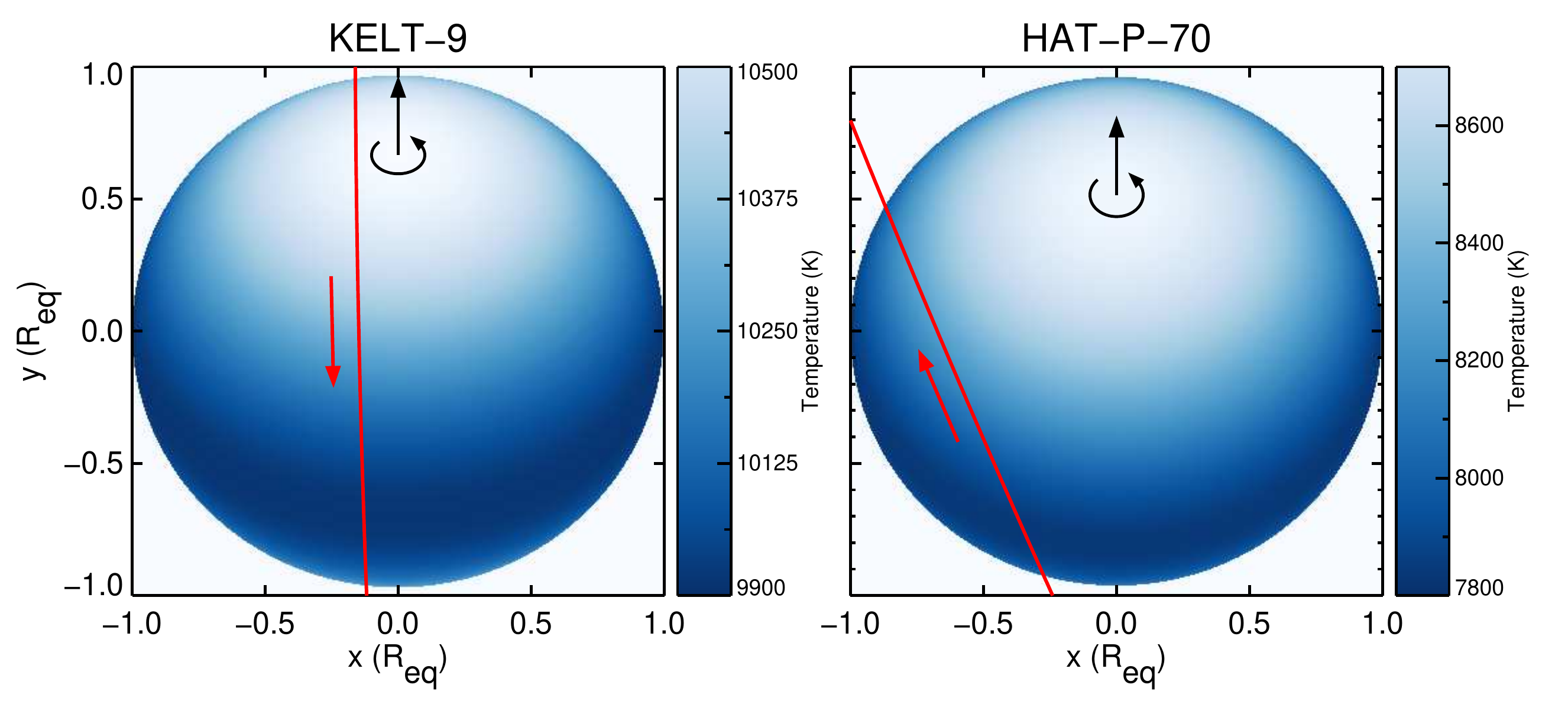}
   \figcaption{Surface temperature maps and transit chords (red lines) for KELT-9 (left) and HAT-P-70 (right). Higher temperatures are indicated by lighter colors. The sky-projected polar and equatorial radii are to-scale relative to each star's equatorial radius. The stellar poles are marked by  vertical black arrows, where the circular arrow shows the direction of stellar rotation. 
\label{fig:tmaps}}

\end{figure*}

For each line we calculate the transmission spectrum at 200 equally spaced points across the transit chords shown in \autoref{fig:tmaps}. We repeat the transmission spectrum calculations for the uniform temperature case. We then shift the transmission spectra into the rest frame of the planet and calculate an average in-transit transmission spectrum which excludes transit points during ingress and egress. We also calculate spectroscopic light curves for each line by integrating the equivalent width of the planetary rest-frame transmission spectrum across 1.5 \AA\ centered on the rest wavelength of the transition.      

\section{RESULTS}
\label{sec:results}

\begin{figure*}[htbp]
   \centering
   \includegraphics[scale=.75,clip,trim=15mm 20mm 15mm 20mm,angle=0]{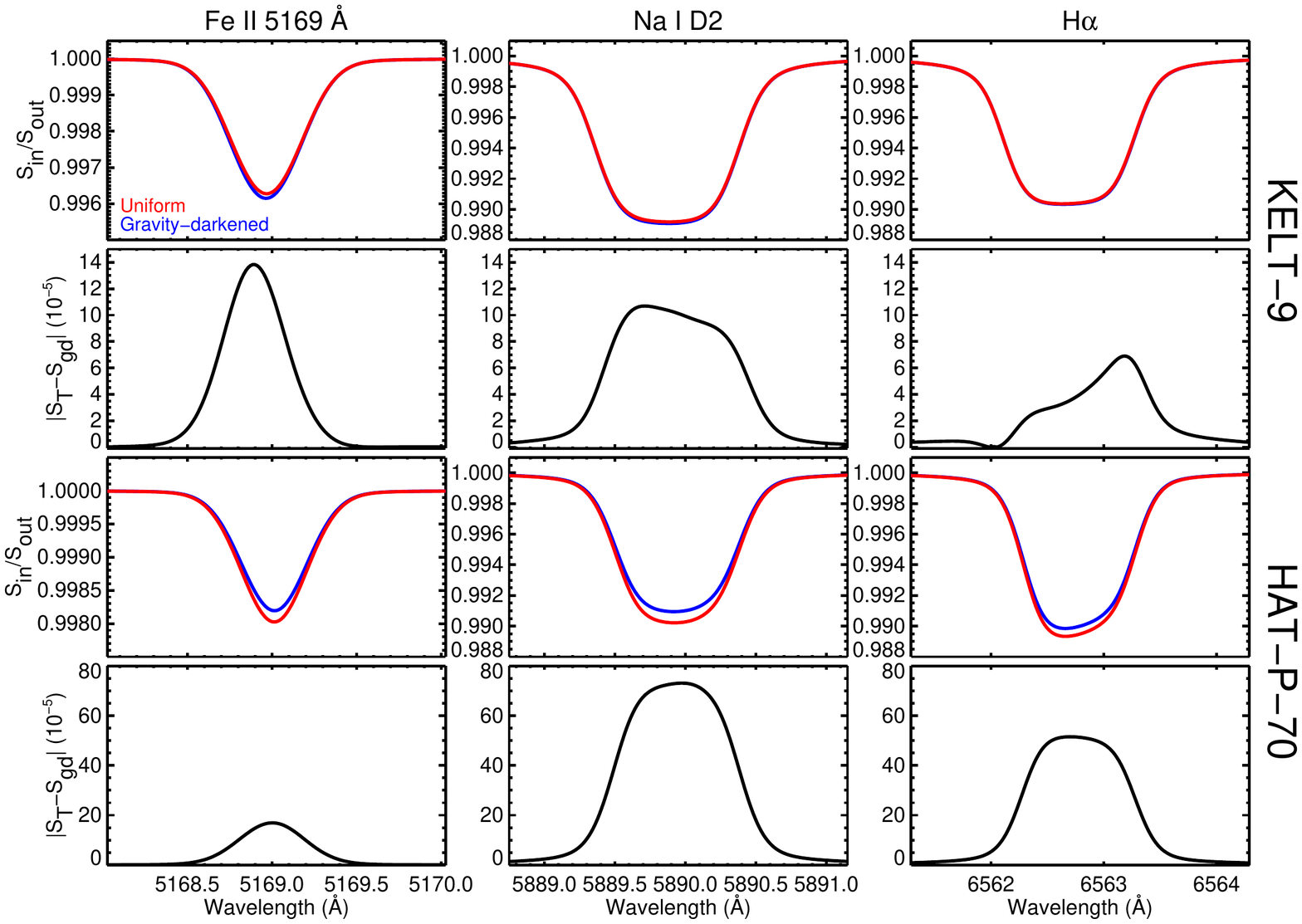}
   \figcaption{Simulated transmission spectra ($S_\text{in}/S_\text{out}$; first and third rows) and the absolute difference between the gravity-darkened and uniform temperature cases ($|S_\text{T}-S_\text{gd}|$; second and fourth rows). The KELT-9 b spectra are in the top two rows and those for HAT-P-70 b are in the bottom two rows. The uniform stellar temperature case is displayed in red and the gravity-darkened case in blue. Note the different y-axis scaling. In all cases the magnitude of the difference in the transmission spectra is less than a part per thousand.  
\label{fig:tspecs}}

\end{figure*}

\begin{figure*}[htbp]
   \centering
   \includegraphics[scale=.75,clip,trim=15mm 20mm 15mm 20mm,angle=0]{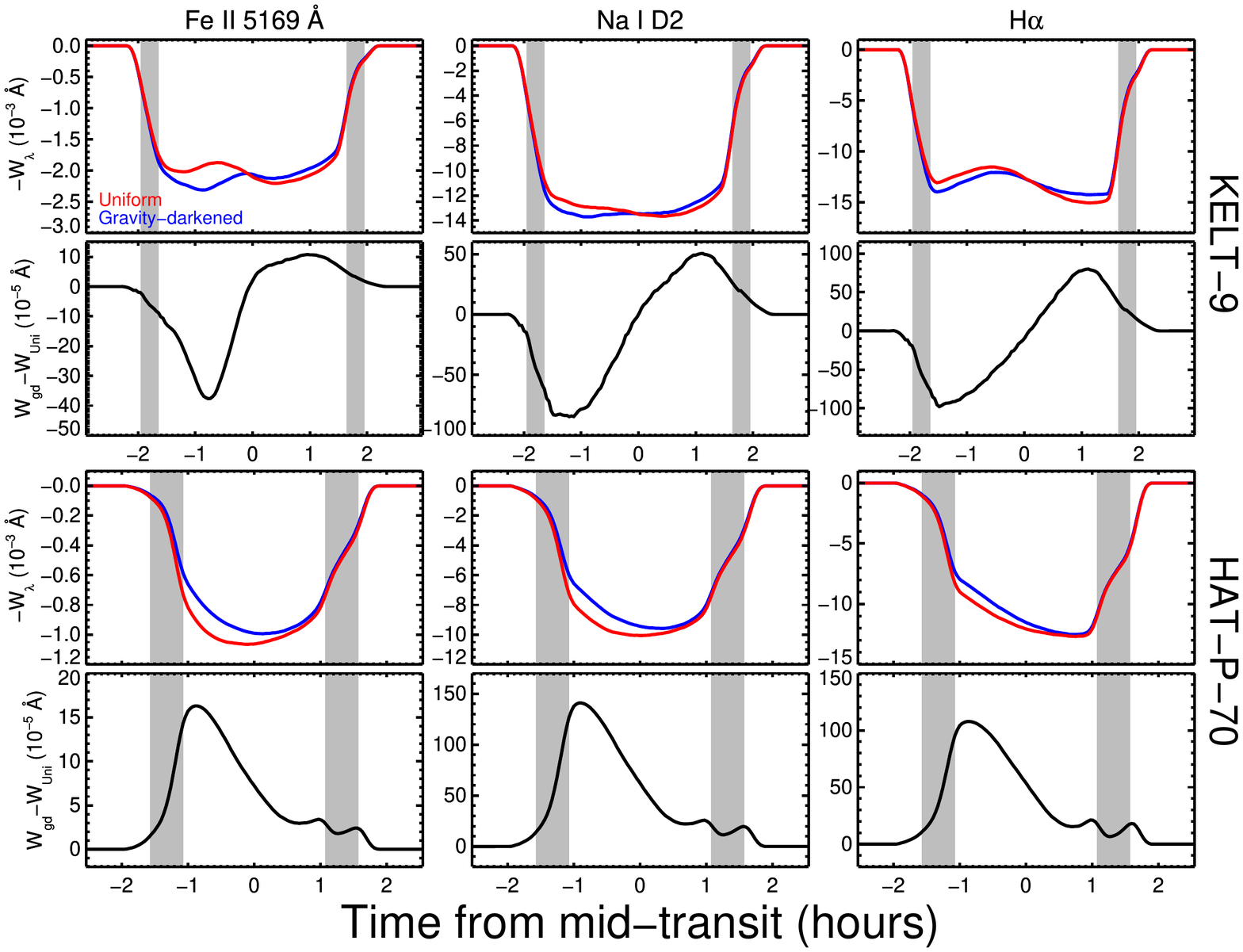}
   \figcaption{Spectroscopic light curves for each atomic transition for a gravity-darkened stellar surface (blue lines) and uniform temperature surface (red lines). The shaded gray regions indicate the ingress and egress portions of the white light transit. The second and fourth rows show the difference in the light curves from the first and third rows, respectively. The light curve differences are larger for HAT-P-70 b, reflecting the fact that the planet transits a cooler chord of the star relative to the single-temperature $T_\text{eff}$ values. Note that the transit duration is slightly longer than the white light transit duration due to early and trailing occultation of the star by the planet's extended atmosphere.
\label{fig:trans}}

\end{figure*}

We show the results of our simulations in \autoref{fig:tspecs} and \autoref{fig:trans} which detail differences in the average transmission spectra and spectroscopic light curves, respectively. In both figures we display the gravity darkened case in blue and the uniform temperature case in red. In the first and third rows of each figure we show the two different cases plotted on top of one another and in the second and fourth rows we plot the difference between the gravity-darkened and uniform temperature simulations.

It is immediately clear from \autoref{fig:tspecs} that the difference between gravity-darkened transmission spectra and those calculated using a uniform temperature stellar surface is small in all cases. For KELT-9 b the difference is $< 10^{-4}$ for all lines; for HAT-P-70 b the differences are larger but still below the level of $\approx 10^{-3}$. There are small changes in line morphology for H$\alpha$ in both systems and a small difference in morphology for \ion{Na}{1} D2 for KELT-9 but the line shape changes are an order of magnitude smaller than the already negligible changes in line depth. We note that any morphology changes between the gravity-darkened and uniform scenarios are too small to result in a detectable velocity shift in the transmission spectrum. Thus wind measurements for hot and ultra-hot Jupiters, which are generally on the order of a few kilometers per second \citep[e.g.,][]{brogi16,seidel19,ehrenreich20,bello21}, are likely unaffected by the assumption of a uniform surface temperature.

Currently, the most sensitive telescope-spectrograph combinations can achieve uncertainties on average optical atomic transmission spectra of $\approx 0.5 - 2.0 \times 10^{-3}$ for hot Jupiters around $V \sim 7-8$ magnitude stars \citep[e.g.,][]{cauley19,yan21,cauley21}. Cross-correlation retrievals of metal species can reach precision levels of $\approx 0.001 - 0.1 \times 10^{-3}$ by leveraging absorption from many lines of single atomic or molecular species \citep[e.g.,][]{hoeijmakers19,ehrenreich20,hoeijmakers20,bello21}, although in this case the interpretation of changes in line depth, i.e., whether or not differences in line depth can be attributed to gravity-darkening, may be obscured by the effective co-adding of dozens or hundreds of lines. We speculate that changes in the cross-correlation signal strength due to gravity-darkening would be even smaller than in the single line case since changes in the more temperature-sensitive transitions will be averaged out by the less temperature-sensitive lines in the cross-correlation analysis. Thus the possible differences induced by gravity-darkening are comparable to the noise levels in the most precise high-resolution transmission spectrum signals presently achievable.  

The differences in the spectroscopic light curves from \autoref{fig:trans} are potentially more interesting. For both objects, which have very different transit chords, the spectroscopic light curves change in both shape and depth for all of the atomic lines. For KELT-9 b the gravity-darkened case shows deeper absorption at the beginning of the transit and shallower absorption near the end which reflects the fact that the planet's transit chord initially crosses the hotter polar region and then the cooler equatorial region. For HAT-P-70 b the transit chord primarily intersects the cooler equatorial region which produces the uniformly weaker spectroscopic light curves for the gravity-darkened case. For both systems the equivalent width differences between the gravity-darkened and uniform temperature curves is $\approx 0.1 - 1.0$ $\times 10^{-3}$. Thus a general rule for determining how the transmission spectrum of a planet will change when gravity-darkening is included is to estimate the difference between the uniform $T_\text{eff}$ value and the average temperature of the stellar surface being occulted. If the occulted region is cooler, the gravity-darkened transit will be shallower; if it is hotter then the gravity-darkened transit will be stronger. Thus the flux contribution from the local stellar surface is more important for the strength of the transmission spectrum than changes in the photospheric line strength due to the stellar temperature gradient.

The difference in shape between the uniform and gravity-darkened case for KELT-9 b is informative concerning the negligible difference in the planet's average transmission spectrum: the transmission spectra during the first half of the transit are deeper in the gravity-darkened case but then are weaker during the second half. The in-transit spectra thus average out to be quite similar to those from the uniform temperature scenario. The same reasoning illustrates why HAT-P-70 b's gravity-darkened spectra are shallower: the entire spectroscopic light curve is shallower thus resulting in weaker average transmission spectra.

Although the differences between the light curves in \autoref{fig:trans} are small, they are approaching the level of the uncertainties for individual equivalent width, or absorption percentage, in hot planet spectroscopic time series measurements. For example, \citet{cauley19} report uncertainties on H$\alpha$ transmission spectrum equivalent widths of $\approx 0.1 - 0.5$ $\times 10^{-3}$ for KELT-9 b. \citet{yan21}, using the HARPS-N and CARMENES instruments, are able to achieve errors on individual H$\alpha$ transmission spectrum equivalent widths of $\approx 1.0 - 3.0$ $\times 10^{-3}$ for WASP-33 b. Assuming similar transmission efficiencies for the next generation of high-resolution optical spectrographs and $\approx 10\times$ the collecting area of a 30-meter telescope versus a 10-meter facility, spectroscopic light curve uncertainties for individual atomic lines could reach the level of $\approx 0.03 - 0.10$ $\times 10^{-3}$. Such precision would allow the uniform and gravity-darkened cases to be distinguished for systems with pole-to-equator temperature differentials similar to those for KELT-9 and HAT-P-70.
\newline

\section{CONCLUSIONS}
\label{sec:conclusion}

We explored how non-uniform stellar surface temperatures, caused by gravity-darkening as a result of high stellar spin rates, can affect the observed atomic transmission spectra of short-period giant planets. We focus on two systems, KELT-9 b and HAT-P-70 b, and find that the magnitude of the gravity-darkening effect is currently comparable to or smaller than the noise levels achievable for high-resolution transmission spectra using state-of-the-art facilities. Although we specifically simulate two systems our results should be widely applicable to any hot and ultra-hot Jupiters transiting gravity-darkened stars. In other words, the magnitude of the gravity-darkening effect should be comparable for most similar systems of interest. Other effects at play in the interpretation and modeling of high-resolution transmission spectra are currently more important considering the present level of achievable precision (e.g., stellar activity, accurate modeling of CLVs and the planetary Doppler shadow, intrinsic atmospheric variability). For the time being we suggest that gravity-darkened stellar surfaces can safely be ignored when modeling and interpreting high-resolution transmission spectra of hot gas giants. However, it is likely that the era of 30-meter telescopes will require gravity-darkening to be taken into account when modeling transmission spectra of planet's orbiting such stars.     

%\begin{acknowledgments}
\bigskip\bigskip

We thank the referee for their careful reading of the manuscript, which resulted in a much improved paper. This research has made use of the NASA Exoplanet Archive, which is operated by the California Institute of Technology, under contract with the National Aeronautics and Space Administration under the Exoplanet Exploration Program. Additionally, this work has made use of NASA's Astrophysical Data System and of the SIMBAD database, which is operated at CDS, Strasbourg, France. This work has also made use of the VALD database, operated at Uppsala University, the Institute of Astronomy RAS in Moscow, and the University of Vienna.

%\end{acknowledgments}

\noindent\software{\texttt{EXOFAST}, \citet{eastman13}; \texttt{Spectroscopy Made Easy (SME)}, \citet{valenti96,piskunov17}}

\bibliography{references}{}

\begin{thebibliography}{}
\expandafter\ifx\csname natexlab\endcsname\relax\def\natexlab#1{#1}\fi
\providecommand{\url}[1]{\href{#1}{#1}}
\providecommand{\dodoi}[1]{doi:~\href{http://doi.org/#1}{\nolinkurl{#1}}}
\providecommand{\doeprint}[1]{\href{http://ascl.net/#1}{\nolinkurl{http://ascl.net/#1}}}
\providecommand{\doarXiv}[1]{\href{https://arxiv.org/abs/#1}{\nolinkurl{https://arxiv.org/abs/#1}}}

\bibitem[{Ahlers {et~al.}(2014)Ahlers, Seubert, \& Barnes}]{ahlers2014spin}
Ahlers, J.~P., Seubert, S.~A., \& Barnes, J.~W. 2014, The Astrophysical
  Journal, 786, 131

\bibitem[{{Ahlers} {et~al.}(2020{\natexlab{a}}){Ahlers}, {Johnson}, {Stassun},
  {Col{\'o}n}, {Barnes}, {Stevens}, {Beatty}, {Gaudi}, {Collins}, {Rodriguez},
  {Ricker}, {Vanderspek}, {Latham}, {Seager}, {Winn}, {Jenkins}, {Caldwell},
  {Goeke}, {Osborn}, {Paegert}, {Rowden}, \& {Tenenbaum}}]{ahlers20}
{Ahlers}, J.~P., {Johnson}, M.~C., {Stassun}, K.~G., {et~al.}
  2020{\natexlab{a}}, \aj, 160, 4, \dodoi{10.3847/1538-3881/ab8fa3}

\bibitem[{{Ahlers} {et~al.}(2020{\natexlab{b}}){Ahlers}, {Kruse}, {Col{\'o}n},
  {Dorval}, {Talens}, {Snellen}, {Albrecht}, {Otten}, {Ricker}, {Vanderspek},
  {Latham}, {Seager}, {Winn}, {Jenkins}, {Haworth}, {Cartwright}, {Morris},
  {Rowden}, {Tenenbaum}, \& {Ting}}]{ahlers20a}
{Ahlers}, J.~P., {Kruse}, E., {Col{\'o}n}, K.~D., {et~al.} 2020{\natexlab{b}},
  \apj, 888, 63, \dodoi{10.3847/1538-4357/ab59d0}

\bibitem[{{Allart} {et~al.}(2018){Allart}, {Bourrier}, {Lovis}, {Ehrenreich},
  {Spake}, {Wyttenbach}, {Pino}, {Pepe}, {Sing}, \& {Lecavelier des
  Etangs}}]{allart18}
{Allart}, R., {Bourrier}, V., {Lovis}, C., {et~al.} 2018, Science, 362, 1384,
  \dodoi{10.1126/science.aat5879}

\bibitem[{{Barnes}(2009)}]{barnes09}
{Barnes}, J.~W. 2009, \apj, 705, 683, \dodoi{10.1088/0004-637X/705/1/683}

\bibitem[{{Barnes} {et~al.}(2011){Barnes}, {Linscott}, \& {Shporer}}]{barnes11}
{Barnes}, J.~W., {Linscott}, E., \& {Shporer}, A. 2011, \apjs, 197, 10,
  \dodoi{10.1088/0067-0049/197/1/10}

\bibitem[{{Bello-Arufe} {et~al.}(2021){Bello-Arufe}, {Cabot}, {Mendon{\c{c}}a},
  {Buchhave}, \& {Rathcke}}]{bello21}
{Bello-Arufe}, A., {Cabot}, S. H.~C., {Mendon{\c{c}}a}, J.~M., {Buchhave},
  L.~A., \& {Rathcke}, A.~D. 2021, arXiv e-prints, arXiv:2112.03292.
\newblock \doarXiv{2112.03292}

\bibitem[{{B{\'e}tr{\'e}mieux}(2016)}]{betremieux16}
{B{\'e}tr{\'e}mieux}, Y. 2016, \mnras, 456, 4051, \dodoi{10.1093/mnras/stv2955}

\bibitem[{{Bourrier} {et~al.}(2013){Bourrier}, {Lecavelier des Etangs},
  {Dupuy}, {Ehrenreich}, {Vidal-Madjar}, {H{\'e}brard}, {Ballester},
  {D{\'e}sert}, {Ferlet}, {Sing}, \& {Wheatley}}]{bourrier13}
{Bourrier}, V., {Lecavelier des Etangs}, A., {Dupuy}, H., {et~al.} 2013, A\&A,
  551, A63, \dodoi{10.1051/0004-6361/201220533}

\bibitem[{{Bourrier} {et~al.}(2018){Bourrier}, {Lecavelier des Etangs},
  {Ehrenreich}, {Sanz-Forcada}, {Allart}, {Ballester}, {Buchhave}, {Cohen},
  {Deming}, {Evans}, {Garc{\'{\i}}a Mu{\~n}oz}, {Henry}, {Kataria}, {Lavvas},
  {Lewis}, {L{\'o}pez-Morales}, {Marley}, {Sing}, \& {Wakeford}}]{bourrier18}
{Bourrier}, V., {Lecavelier des Etangs}, A., {Ehrenreich}, D., {et~al.} 2018,
  A\&A, 620, A147, \dodoi{10.1051/0004-6361/201833675}

\bibitem[{{Brogi} {et~al.}(2016){Brogi}, {de Kok}, {Albrecht}, {Snellen},
  {Birkby}, \& {Schwarz}}]{brogi16}
{Brogi}, M., {de Kok}, R.~J., {Albrecht}, S., {et~al.} 2016, ApJ, 817, 106,
  \dodoi{10.3847/0004-637X/817/2/106}

\bibitem[{{Casasayas-Barris} {et~al.}(2017){Casasayas-Barris}, {Palle},
  {Nowak}, {Yan}, {Nortmann}, \& {Murgas}}]{casasayas17}
{Casasayas-Barris}, N., {Palle}, E., {Nowak}, G., {et~al.} 2017, \aap, 608,
  A135, \dodoi{10.1051/0004-6361/201731956}

\bibitem[{{Casasayas-Barris} {et~al.}(2019){Casasayas-Barris}, {Pall{\'e}},
  {Yan}, {Chen}, {Kohl}, {Stangret}, {Parviainen}, {Helling}, {Watanabe},
  {Czesla}, {Fukui}, {Monta{\~n}{\'e}s-Rodr{\'\i}guez}, {Nagel}, {Narita},
  {Nortmann}, {Nowak}, {Schmitt}, \& {Zapatero Osorio}}]{casasayas19}
{Casasayas-Barris}, N., {Pall{\'e}}, E., {Yan}, F., {et~al.} 2019, \aap, 628,
  A9, \dodoi{10.1051/0004-6361/201935623}

\bibitem[{{Cauley} {et~al.}(2018){Cauley}, {Kuckein}, {Redfield}, {Shkolnik},
  {Denker}, {Llama}, \& {Verma}}]{cauley18}
{Cauley}, P.~W., {Kuckein}, C., {Redfield}, S., {et~al.} 2018, AJ, 156, 189,
  \dodoi{10.3847/1538-3881/aaddf9}

\bibitem[{{Cauley} {et~al.}(2019){Cauley}, {Shkolnik}, {Ilyin}, {Strassmeier},
  {Redfield}, \& {Jensen}}]{cauley19}
{Cauley}, P.~W., {Shkolnik}, E.~L., {Ilyin}, I., {et~al.} 2019, AJ, 157, 69,
  \dodoi{10.3847/1538-3881/aaf725}

\bibitem[{{Cauley} {et~al.}(2021){Cauley}, {Wang}, {Shkolnik}, {Ilyin},
  {Strassmeier}, {Redfield}, \& {Jensen}}]{cauley21}
{Cauley}, P.~W., {Wang}, J., {Shkolnik}, E.~L., {et~al.} 2021, AJ, 161, 152,
  \dodoi{10.3847/1538-3881/abde43}

\bibitem[{{Czesla} {et~al.}(2015){Czesla}, {Klocov{\'a}}, {Khalafinejad},
  {Wolter}, \& {Schmitt}}]{czesla15}
{Czesla}, S., {Klocov{\'a}}, T., {Khalafinejad}, S., {Wolter}, U., \&
  {Schmitt}, J.~H.~M.~M. 2015, A\&A, 582, A51,
  \dodoi{10.1051/0004-6361/201526386}

\bibitem[{{Deming} {et~al.}(2013){Deming}, {Wilkins}, {McCullough}, {Burrows},
  {Fortney}, {Agol}, {Dobbs-Dixon}, {Madhusudhan}, {Crouzet}, {Desert},
  {Gilliland}, {Haynes}, {Knutson}, {Line}, {Magic}, {Mandell}, {Ranjan},
  {Charbonneau}, {Clampin}, {Seager}, \& {Showman}}]{deming13}
{Deming}, D., {Wilkins}, A., {McCullough}, P., {et~al.} 2013, \apj, 774, 95,
  \dodoi{10.1088/0004-637X/774/2/95}

\bibitem[{{Eastman} {et~al.}(2013){Eastman}, {Gaudi}, \& {Agol}}]{eastman13}
{Eastman}, J., {Gaudi}, B.~S., \& {Agol}, E. 2013, \pasp, 125, 83,
  \dodoi{10.1086/669497}

\bibitem[{{Ehrenreich} {et~al.}(2015){Ehrenreich}, {Bourrier}, {Wheatley},
  {Lecavelier des Etangs}, {H{\'e}brard}, {Udry}, {Bonfils}, {Delfosse},
  {D{\'e}sert}, {Sing}, \& {Vidal-Madjar}}]{ehrenreich15}
{Ehrenreich}, D., {Bourrier}, V., {Wheatley}, P.~J., {et~al.} 2015, Nature,
  522, 459, \dodoi{10.1038/nature14501}

\bibitem[{{Ehrenreich} {et~al.}(2020){Ehrenreich}, {Lovis}, {Allart}, {Zapatero
  Osorio}, {Pepe}, {Cristiani}, {Rebolo}, {Santos}, {Borsa}, {Demangeon},
  {Dumusque}, {Gonz{\'a}lez Hern{\'a}ndez}, {Casasayas-Barris},
  {S{\'e}gransan}, {Sousa}, {Abreu}, {Adibekyan}, {Affolter}, {Allende Prieto},
  {Alibert}, {Aliverti}, {Alves}, {Amate}, {Avila}, {Baldini}, {Bandy}, {Benz},
  {Bianco}, {Bolmont}, {Bouchy}, {Bourrier}, {Broeg}, {Cabral}, {Calderone},
  {Pall{\'e}}, {Cegla}, {Cirami}, {Coelho}, {Conconi}, {Coretti}, {Cumani},
  {Cupani}, {Dekker}, {Delabre}, {Deiries}, {D'Odorico}, {Di Marcantonio},
  {Figueira}, {Fragoso}, {Genolet}, {Genoni}, {G{\'e}nova Santos}, {Hara},
  {Hughes}, {Iwert}, {Kerber}, {Knudstrup}, {Land oni}, {Lavie}, {Lizon},
  {Lendl}, {Lo Curto}, {Maire}, {Manescau}, {Martins}, {M{\'e}gevand },
  {Mehner}, {Micela}, {Modigliani}, {Molaro}, {Monteiro}, {Monteiro},
  {Moschetti}, {M{\"u}ller}, {Nunes}, {Oggioni}, {Oliveira}, {Pariani},
  {Pasquini}, {Poretti}, {Rasilla}, {Redaelli}, {Riva}, {Santana Tschudi},
  {Santin}, {Santos}, {Segovia Milla}, {Seidel}, {Sosnowska}, {Sozzetti},
  {Span{\`o}}, {Su{\'a}rez Mascare{\~n}o}, {Tabernero}, {Tenegi}, {Udry},
  {Zanutta}, \& {Zerbi}}]{ehrenreich20}
{Ehrenreich}, D., {Lovis}, C., {Allart}, R., {et~al.} 2020, \nat, 580, 597,
  \dodoi{10.1038/s41586-020-2107-1}

\bibitem[{{Espinosa Lara} \& {Rieutord}(2011)}]{espinosalara2011}
{Espinosa Lara}, F., \& {Rieutord}, M. 2011, \aap, 533, A43,
  \dodoi{10.1051/0004-6361/201117252}

\bibitem[{{Garai} {et~al.}(2021){Garai}, {Pribulla}, {Parviainen}, {Pall{\'e}},
  {Claret}, {Szigeti}, {B{\'e}jar}, {Casasayas-Barris}, {Crouzet}, {Fukui},
  {Chen}, {Kawauchi}, {Klagyivik}, {Kurita}, {Kusakabe}, {de Leon},
  {Livingston}, {Luque}, {Mori}, {Murgas}, {Narita}, {Nishiumi}, {Oshagh},
  {Szab{\'o}}, {Tamura}, {Terada}, \& {Watanabe}}]{garai21}
{Garai}, Z., {Pribulla}, T., {Parviainen}, H., {et~al.} 2021, \mnras, 508,
  5514, \dodoi{10.1093/mnras/stab2929}

\bibitem[{{Gaudi} {et~al.}(2017){Gaudi}, {Stassun}, {Collins}, {Beatty},
  {Zhou}, {Latham}, {Bieryla}, {Eastman}, {Siverd}, {Crepp}, {Gonzales},
  {Stevens}, {Buchhave}, {Pepper}, {Johnson}, {Colon}, {Jensen}, {Rodriguez},
  {Bozza}, {Novati}, {D'Ago}, {Dumont}, {Ellis}, {Gaillard}, {Jang-Condell},
  {Kasper}, {Fukui}, {Gregorio}, {Ito}, {Kielkopf}, {Manner}, {Matt}, {Narita},
  {Oberst}, {Reed}, {Scarpetta}, {Stephens}, {Yeigh}, {Zambelli}, {Fulton},
  {Howard}, {James}, {Penny}, {Bayliss}, {Curtis}, {Depoy}, {Esquerdo},
  {Gould}, {Joner}, {Kuhn}, {Labadie-Bartz}, {Lund}, {Marshall}, {McLeod},
  {Pogge}, {Relles}, {Stockdale}, {Tan}, {Trueblood}, \& {Trueblood}}]{gaudi17}
{Gaudi}, B.~S., {Stassun}, K.~G., {Collins}, K.~A., {et~al.} 2017, Nature, 546,
  514, \dodoi{10.1038/nature22392}

\bibitem[{{Herman} {et~al.}(2018){Herman}, {de Mooij}, {Huang}, \&
  {Jayawardhana}}]{herman18}
{Herman}, M.~K., {de Mooij}, E. J.~W., {Huang}, C.~X., \& {Jayawardhana}, R.
  2018, \aj, 155, 13, \dodoi{10.3847/1538-3881/aa991f}

\bibitem[{{Hoeijmakers} {et~al.}(2019){Hoeijmakers}, {Ehrenreich}, {Kitzmann},
  {Allart}, {Grimm}, {Seidel}, {Wyttenbach}, {Pino}, {Nielsen}, {Fisher},
  {Rimmer}, {Bourrier}, {Cegla}, {Lavie}, {Lovis}, {Patzer}, {Stock}, {Pepe},
  \& {Heng}}]{hoeijmakers19}
{Hoeijmakers}, H.~J., {Ehrenreich}, D., {Kitzmann}, D., {et~al.} 2019, A\&A,
  627, A165, \dodoi{10.1051/0004-6361/201935089}

\bibitem[{{Hoeijmakers} {et~al.}(2020){Hoeijmakers}, {Cabot}, {Zhao},
  {Buchhave}, {Tronsgaard}, {Davis}, {Kitzmann}, {Grimm}, {Cegla}, {Bourrier},
  {Ehrenreich}, {Heng}, {Lovis}, \& {Fischer}}]{hoeijmakers20}
{Hoeijmakers}, H.~J., {Cabot}, S. H.~C., {Zhao}, L., {et~al.} 2020, \aap, 641,
  A120, \dodoi{10.1051/0004-6361/202037437}

\bibitem[{{Johnson} {et~al.}(2018){Johnson}, {Rodriguez}, {Zhou}, {Gonzales},
  {Cargile}, {Crepp}, {Penev}, {Stassun}, {Gaudi}, {Col{\'o}n}, {Stevens},
  {Strassmeier}, {Ilyin}, {Collins}, {Kielkopf}, {Oberst}, {Maritch}, {Reed},
  {Gregorio}, {Bozza}, {Calchi Novati}, {D'Ago}, {Scarpetta}, {Zambelli},
  {Latham}, {Bieryla}, {Cochran}, {Endl}, {Tayar}, {Serenelli}, {Silva
  Aguirre}, {Clarke}, {Martinez}, {Spencer}, {Trump}, {Joner}, {Bugg}, {Hintz},
  {Stephens}, {Arredondo}, {Benzaid}, {Yazdi}, {McLeod}, {Jensen}, {Hancock},
  {Sorber}, {Kasper}, {Jang-Condell}, {Beatty}, {Carroll}, {Eastman}, {James},
  {Kuhn}, {Labadie-Bartz}, {Lund}, {Mallonn}, {Pepper}, {Siverd}, {Yao},
  {Cohen}, {Curtis}, {DePoy}, {Fulton}, {Penny}, {Relles}, {Stockdale}, {Tan},
  \& {Villanueva}}]{johnson18}
{Johnson}, M.~C., {Rodriguez}, J.~E., {Zhou}, G., {et~al.} 2018, \aj, 155, 100,
  \dodoi{10.3847/1538-3881/aaa5af}

\bibitem[{{Kreidberg} {et~al.}(2014){Kreidberg}, {Bean}, {D{\'e}sert},
  {Benneke}, {Deming}, {Stevenson}, {Seager}, {Berta-Thompson}, {Seifahrt}, \&
  {Homeier}}]{kreidberg14}
{Kreidberg}, L., {Bean}, J.~L., {D{\'e}sert}, J.-M., {et~al.} 2014, \nat, 505,
  69, \dodoi{10.1038/nature12888}

\bibitem[{{Masuda}(2015)}]{masuda15}
{Masuda}, K. 2015, \apj, 805, 28, \dodoi{10.1088/0004-637X/805/1/28}

\bibitem[{{Nugroho} {et~al.}(2020){Nugroho}, {Gibson}, {de Mooij}, {Watson},
  {Kawahara}, \& {Merritt}}]{nugroho20a}
{Nugroho}, S.~K., {Gibson}, N.~P., {de Mooij}, E. J.~W., {et~al.} 2020, \mnras,
  496, 504, \dodoi{10.1093/mnras/staa1459}

\bibitem[{{Piskunov} \& {Valenti}(2017)}]{piskunov17}
{Piskunov}, N., \& {Valenti}, J.~A. 2017, \aap, 597, A16,
  \dodoi{10.1051/0004-6361/201629124}

\bibitem[{{Rackham} {et~al.}(2018){Rackham}, {Apai}, \& {Giampapa}}]{rackham18}
{Rackham}, B.~V., {Apai}, D., \& {Giampapa}, M.~S. 2018, ApJ, 853, 122,
  \dodoi{10.3847/1538-4357/aaa08c}

\bibitem[{{Rainer} {et~al.}(2021){Rainer}, {Borsa}, {Pino}, {Frustagli},
  {Brogi}, {Biazzo}, {Bonomo}, {Carleo}, {Claudi}, {Gratton}, {Lanza},
  {Maggio}, {Maldonado}, {Mancini}, {Micela}, {Scandariato}, {Sozzetti},
  {Buchschacher}, {Cosentino}, {Covino}, {Ghedina}, {Gonzalez}, {Leto}, {Lodi},
  {Martinez Fiorenzano}, {Molinari}, {Molinaro}, {Nardiello}, {Oliva},
  {Pagano}, {Pedani}, {Piotto}, \& {Poretti}}]{rainer21}
{Rainer}, M., {Borsa}, F., {Pino}, L., {et~al.} 2021, A\&A, 649, A29,
  \dodoi{10.1051/0004-6361/202039247}

\bibitem[{{Redfield} {et~al.}(2008){Redfield}, {Endl}, {Cochran}, \&
  {Koesterke}}]{redfield08}
{Redfield}, S., {Endl}, M., {Cochran}, W.~D., \& {Koesterke}, L. 2008, ApJ,
  673, L87, \dodoi{10.1086/527475}

\bibitem[{{Robinson} {et~al.}(2017){Robinson}, {Fortney}, \&
  {Hubbard}}]{robinson17}
{Robinson}, T.~D., {Fortney}, J.~J., \& {Hubbard}, W.~B. 2017, \apj, 850, 128,
  \dodoi{10.3847/1538-4357/aa951e}

\bibitem[{{Salz} {et~al.}(2018){Salz}, {Czesla}, {Schneider}, {Nagel},
  {Schmitt}, {Nortmann}, {Alonso-Floriano}, {L{\'o}pez-Puertas}, {Lamp{\'o}n},
  {Bauer}, {Snellen}, {Pall{\'e}}, {Caballero}, {Yan}, {Chen}, {Sanz-Forcada},
  {Amado}, {Quirrenbach}, {Ribas}, {Reiners}, {B{\'e}jar}, {Casasayas-Barris},
  {Cort{\'e}s-Contreras}, {Dreizler}, {Guenther}, {Henning}, {Jeffers},
  {Kaminski}, {K{\"u}rster}, {Lafarga}, {Lara}, {Molaverdikhani}, {Montes},
  {Morales}, {S{\'a}nchez-L{\'o}pez}, {Seifert}, {Zapatero Osorio}, \&
  {Zechmeister}}]{salz18}
{Salz}, M., {Czesla}, S., {Schneider}, P.~C., {et~al.} 2018, \aap, 620, A97,
  \dodoi{10.1051/0004-6361/201833694}

\bibitem[{{Seidel} {et~al.}(2019){Seidel}, {Ehrenreich}, {Pino}, {Bourrier},
  {Lavie}, {Allart}, {Wyttenbach}, \& {Lovis}}]{seidel19}
{Seidel}, J.~V., {Ehrenreich}, D., {Pino}, L., {et~al.} 2019, arXiv e-prints,
  arXiv:1912.02787.
\newblock \doarXiv{1912.02787}

\bibitem[{{Seidel} {et~al.}(2020){Seidel}, {Ehrenreich}, {Bourrier}, {Allart},
  {Attia}, {Hoeijmakers}, {Lendl}, {Linder}, {Wyttenbach}, {Astudillo-Defru},
  {Bayliss}, {Cegla}, {Heng}, {Lavie}, {Lovis}, {Melo}, {Pepe}, {dos Santos},
  {S{\'e}gransan}, \& {Udry}}]{seidel20a}
{Seidel}, J.~V., {Ehrenreich}, D., {Bourrier}, V., {et~al.} 2020, \aap, 641,
  L7, \dodoi{10.1051/0004-6361/202038497}

\bibitem[{{Sing} {et~al.}(2016){Sing}, {Fortney}, {Nikolov}, {Wakeford},
  {Kataria}, {Evans}, {Aigrain}, {Ballester}, {Burrows}, {Deming},
  {D{\'e}sert}, {Gibson}, {Henry}, {Huitson}, {Knutson}, {Lecavelier Des
  Etangs}, {Pont}, {Showman}, {Vidal-Madjar}, {Williamson}, \&
  {Wilson}}]{sing16}
{Sing}, D.~K., {Fortney}, J.~J., {Nikolov}, N., {et~al.} 2016, Nature, 529, 59,
  \dodoi{10.1038/nature16068}

\bibitem[{{Stangret} {et~al.}(2020){Stangret}, {Casasayas-Barris}, {Pall{\'e}},
  {Yan}, {S{\'a}nchez-L{\'o}pez}, \& {L{\'o}pez-Puertas}}]{stangret20}
{Stangret}, M., {Casasayas-Barris}, N., {Pall{\'e}}, E., {et~al.} 2020, \aap,
  638, A26, \dodoi{10.1051/0004-6361/202037541}

\bibitem[{{Szab{\'o}} {et~al.}(2020){Szab{\'o}}, {Pribulla}, {P{\'a}l},
  {B{\'o}di}, {Kiss}, \& {Derekas}}]{szabo20}
{Szab{\'o}}, G.~M., {Pribulla}, T., {P{\'a}l}, A., {et~al.} 2020, \mnras, 492,
  L17, \dodoi{10.1093/mnrasl/slz177}

\bibitem[{{Valenti} \& {Piskunov}(1996)}]{valenti96}
{Valenti}, J.~A., \& {Piskunov}, N. 1996, \aaps, 118, 595

\bibitem[{Von~Zeipel(1924)}]{von1924radiative}
Von~Zeipel, H. 1924, Monthly Notices of the Royal Astronomical Society, 84, 684

\bibitem[{{Watanabe} {et~al.}(2020){Watanabe}, {Narita}, \&
  {Johnson}}]{watanabe20}
{Watanabe}, N., {Narita}, N., \& {Johnson}, M.~C. 2020, \pasj, 72, 19,
  \dodoi{10.1093/pasj/psz140}

\bibitem[{{Weaver} {et~al.}(2021){Weaver}, {L{\'o}pez-Morales}, {Alam},
  {Espinoza}, {Rackham}, {Goyal}, {MacDonald}, {Lewis}, {Apai}, {Bixel},
  {Jord{\'a}n}, {Kirk}, {McGruder}, \& {Osip}}]{weaver21}
{Weaver}, I.~C., {L{\'o}pez-Morales}, M., {Alam}, M.~K., {et~al.} 2021, AJ,
  161, 278, \dodoi{10.3847/1538-3881/abf652}

\bibitem[{{Wyttenbach} {et~al.}(2015){Wyttenbach}, {Ehrenreich}, {Lovis},
  {Udry}, \& {Pepe}}]{wyttenbach15}
{Wyttenbach}, A., {Ehrenreich}, D., {Lovis}, C., {Udry}, S., \& {Pepe}, F.
  2015, A\&A, 577, A62, \dodoi{10.1051/0004-6361/201525729}

\bibitem[{{Yan} {et~al.}(2015){Yan}, {Fosbury}, {Petr-Gotzens}, {Zhao}, \&
  {Pall{\'e}}}]{yan15}
{Yan}, F., {Fosbury}, R.~A.~E., {Petr-Gotzens}, M.~G., {Zhao}, G., \&
  {Pall{\'e}}, E. 2015, \aap, 574, A94, \dodoi{10.1051/0004-6361/201425220}

\bibitem[{{Yan} \& {Henning}(2018)}]{yan18}
{Yan}, F., \& {Henning}, T. 2018, Nature Astronomy, 2, 714,
  \dodoi{10.1038/s41550-018-0503-3}

\bibitem[{{Yan} {et~al.}(2017){Yan}, {Pall{\'e}}, {Fosbury}, {Petr-Gotzens}, \&
  {Henning}}]{yan17}
{Yan}, F., {Pall{\'e}}, E., {Fosbury}, R.~A.~E., {Petr-Gotzens}, M.~G., \&
  {Henning}, T. 2017, \aap, 603, A73, \dodoi{10.1051/0004-6361/201630144}

\bibitem[{{Yan} {et~al.}(2021){Yan}, {Wyttenbach}, {Casasayas-Barris},
  {Reiners}, {Pall{\'e}}, {Henning}, {Molli{\`e}re}, {Czesla}, {Nortmann},
  {Molaverdikhani}, {Chen}, {Snellen}, {Zechmeister}, {Huang}, {Ribas},
  {Quirrenbach}, {Caballero}, {Amado}, {Cont}, {Khalafinejad}, {Khaimova},
  {L{\'o}pez-Puertas}, {Montes}, {Nagel}, {Oshagh}, {Pedraz}, \&
  {Stangret}}]{yan21}
{Yan}, F., {Wyttenbach}, A., {Casasayas-Barris}, N., {et~al.} 2021, \aap, 645,
  A22, \dodoi{10.1051/0004-6361/202039302}

\bibitem[{{Zhang} {et~al.}(2021){Zhang}, {Knutson}, {Wang}, {Dai}, {dos
  Santos}, {Fossati}, {Henry}, {Ehrenreich}, {Alibert}, {Hoyer}, {Wilson}, \&
  {Bonfanti}}]{zhang21}
{Zhang}, M., {Knutson}, H.~A., {Wang}, L., {et~al.} 2021, arXiv e-prints,
  arXiv:2106.05273.
\newblock \doarXiv{2106.05273}

\bibitem[{{Zhou} {et~al.}(2019){Zhou}, {Huang}, {Bakos}, {Hartman}, {Latham},
  {Quinn}, {Collins}, {Winn}, {Wong}, {Kov{\'a}cs}, {Csubry}, {Bhatti},
  {Penev}, {Bieryla}, {Esquerdo}, {Berlind}, {Calkins}, {de Val-Borro},
  {Noyes}, {L{\'a}z{\'a}r}, {Papp}, {S{\'a}ri}, {Kov{\'a}cs}, {Buchhave},
  {Szklenar}, {B{\'e}ky}, {Johnson}, {Cochran}, {Kniazev}, {Stassun}, {Fulton},
  {Shporer}, {Espinoza}, {Bayliss}, {Everett}, {Howell}, {Hellier}, {Anderson},
  {Collier Cameron}, {West}, {Brown}, {Schanche}, {Barkaoui}, {Pozuelos},
  {Gillon}, {Jehin}, {Benkhaldoun}, {Daassou}, {Ricker}, {Vanderspek},
  {Seager}, {Jenkins}, {Lissauer}, {Armstrong}, {Collins}, {Gan}, {Hart},
  {Horne}, {Kielkopf}, {Nielsen}, {Nishiumi}, {Narita}, {Palle}, {Relles},
  {Sefako}, {Tan}, {Davies}, {Goeke}, {Guerrero}, {Haworth}, \&
  {Villanueva}}]{zhou19}
{Zhou}, G., {Huang}, C.~X., {Bakos}, G.~{\'A}., {et~al.} 2019, \aj, 158, 141,
  \dodoi{10.3847/1538-3881/ab36b5}

\end{thebibliography}
\bibliographystyle{aasjournal}

\end{document}